\documentstyle[preprint,prc,aps,epsf]{revtex}
  \def\g{{\gamma}}

  \def\gsl#1{\rlap{\slash}#1} 
   
  \def\p{\gsl p}

  \def\qq{\langle\bar qq\rangle} 
   
  \def\qGq{\langle g\bar q\sigma_{\mu\nu}G^{\mu\nu}q\rangle} 
  \def\GG{\langle{\alpha_s\over\pi}G_{\mu\nu}G^{\mu\nu}\rangle}
  \def\<>#1{\langle#1\rangle} 
  \def\><{\rangle\langle} 
  \def\vec#1{\mbox{\boldmath$#1$}}
\title{Coupled QCD sum rules for positive and negative-parity nucleons}
\author{Yoshihiko Kondo\footnote{kondo@kokugakuin.ac.jp}}
\address{Kokugakuin University, Higashi, Shibuya, Tokyo 150-8440, Japan}
\author{Osamu Morimatsu\footnote{osamu.morimatsu@kek.jp} and
Tetsuo Nishikawa\footnote{nishi@post.kek.jp}}
\address{Institute of Particle and Nuclear Studies, High Energy Accelerator Research Organization, Tukuba, Ibaragi 305-0801, Japan}
\preprint{KEK-TH-1010}

\begin{document}
\draft
\maketitle

\begin{abstract}
A new approach of the QCD sum rule is proposed in which positive and negative-parity baryons couple with each other.
With positive and negative-parity states explicitly taken into account, sum rules are derived by means of the dispersion relation in energy.
The method is applied to the nucleon channel and the parity splitting of the nucleon resonance states is studied.
It is found that the obtained sum rules have a very good Borel stability.
This suggests that the ansatz for the spectral function in the present sum rule approximates the physical spectrum better than the usual lowest pole plus continuum ansatz.
The predicted masses of the positive and negative nucleons reproduce the experimental ones fairly well. 
Especially, the mass difference is extremely close to the experimental value.
\end{abstract}


\newpage
\section{Introduction}

The QCD sum rule invented by Shifman, Vainshtein, and Zakharov~\cite{SVZ} provides us with a way to relate hadron properties to vacuum condensates of quark-gluon composite fields which characterize the structure of QCD vacuum.
The first very successful application of the QCD sum rule has been in mesonic channels~\cite{SVZ,RRY}.
Especially for the vector meson states, the predictions of the masses and decay constants are extremely good.
Then the application has been extended to baryonic channels~\cite{ioffe,chungI}.
The approach has been improved and expanded to various baryons~\cite{BandI,chungII,RRYII,espriu,jido}.
In baryon channels, however, the application has not been so successful as in meson channels.

In the application of the QCD sum rule, the time-ordered correlation function,
\begin{eqnarray}
\Pi(p)=-i\int{d^4x}e^{ipx}\langle0|T(\eta(x)\bar\eta(0))|0\rangle,
\end{eqnarray}
is considered, where $\eta$ is the baryon interpolating field.
The interpolating field whose intrinsic parity is positive couples not only to the positive-parity baryon states but also to the negative-parity baryon states.
Therefore, the contamination from the negative-parity baryons might be the cause of the unsuccessfulness of the QCD sum rule in the baryon channel.
In Ref.~\cite{chungII} Chung et al.\ studied the property of the negative-parity nucleon as well as the positive-parity nucleon. 
They constructed the interpolating fields without derivatives which couple strongly to the positive and negative-parity nucleon, respectively, through the positivity condition of the spectral functions.
Their sum rule, however, does not show a stable plateau which is a necessary condition for the sum rule to be reliable.
In Ref.~\cite{jido} Jido et al.\ proposed to use the \lq\lq old-fashioned'' correlation function,
\begin{eqnarray}
\Pi_{old}(p)=-i\int{d^4x}e^{ipx}\theta(x_0)\langle0|\eta(x)\bar\eta(0)|0\rangle,
\end{eqnarray}
in order to separate the terms of negative-parity baryons from those of positive-parity baryons by operating the projection operators.
They constructed sum rules for positive and negative-parity baryons, respectively, in which they claim to obtain the operator product expansion (OPE) of the \lq\lq old-fashioned'' correlation function from the time-ordered correlation function.
We, however, believe that the OPE of the \lq\lq old-fashioned'' correlation function cannot be obtained from that of the time-ordered correlation function as we will explain.
As long as one uses the OPE of the time-ordered correlation function, the positive and negative baryon states inevitably couple with each other.
Therefore, in our opinion their sum rules do not correctly take into account the coupling of the positive and negative-parity states.

In this paper we propose a new approach of the sum rule in which positive and negative-parity baryons couple with each other.
In the approach we use the dispersion relation in the variable of energy and derive sum rules.
We then apply the method to the nucleon channel and investigate the parity splitting between the positive and negative nucleon states.
We will show that the prediction of the present sum rule is much better than that of the previous works not only for the positive-parity state but also for the negative-parity state.
This suggests that the coupling of the positive and negative-parity states is important in the application of the QCD sum rule to baryon channels.
It is found that once this coupling is taken into account the application of the QCD sum rule for the nucleon can be as successful as for the vector meson.
The paper is organized as follows.
In Sec.~{II} we explain the derivation of the sum rule in which the positive and negative-parity baryons couple with each other.
We then present the results of the application of the sum rule to the nucleon in Sec.~{III}. 
Finally, we summarize the paper in Sec.~{IV}.

\section{QCD sum rule with parity coupling}

The interpolating field of the baryon, $\eta$, is usually constructed as the product of three quark fields, $:qqq:$.
The field, $\eta$, couples to positive and negative-parity resonance states, $|B^n_+\rangle$ and $|B^n_-\rangle$, respectively, as
\begin{eqnarray}
\langle0|\eta(x)|B^n_+(p,s)\rangle&=&\lambda_+^n u_{B^n_+}(p,s)e^{-ipx},\cr
\langle0|\eta(x)|B^n_-(p,s)\rangle&=&\lambda_-^n \gamma_5u_{B^n_-}(p,s)e^{-ipx},
\end{eqnarray}
where $u_B(p,s)$ is a positive energy solution of the free Dirac equation of the baryon~\cite{chungII}.
Therefore, in the zero-width approximation the correlation function can be expressed as
\begin{eqnarray}\label{TimePi}
\Pi(p)&=&\sum_n\left\{|\lambda^n_+|^2{\p+m^n_+\over p^2-{m^n_+}^2}
+|\lambda^n_-|^2{\p-m^n_-\over p^2-{m^n_-}^2}\right\},
\end{eqnarray}
where $m^n_\pm$ is the mass of the $n$-th resonance~\cite{jido}.
Then, we obtain the spectral function, $\rho(p_0)=-{\rm Im}\Pi(p_0+i\epsilon)/\pi$, in the rest frame, $\vec p=0$, as
\begin{eqnarray}\label{Spectral}
\rho(p_0)&=&
P_+\sum_n\left\{|\lambda^n_+|^2\delta(p_0-m^n_+)+|\lambda^n_-|^2\delta(p_0+m^n_-)\right\}
\cr&&+P_-\sum_n\left\{|\lambda^n_+|^2\delta(p_0+m^n_+)+|\lambda^n_-|^2\delta(p_0-m^n_-)\right\},
\end{eqnarray}
where $P_\pm=(\g_0\pm1)/2$. Strictly speaking, $\rho$ is the imaginary part of the retarded correlation function since we approach the real energy axis from the above in the complex energy plain when we take the imaginary part of the correlation function.

One can obtain the QCD sum rule by using the analyticity of the correlation function and replacing the correlation function, $\Pi$, by that in the OPE, $\Pi^{\rm OPE}$ in the deep Euclid region, $p_0^2\rightarrow-\infty$, as
\begin{eqnarray}\label{QSR}
\int_{-\infty}^{\infty}dp_0\rho^{\rm OPE}(p_0)W(p_0)
&=&\int_{-\infty}^{\infty}dp_0\rho(p_0)W(p_0).
\end{eqnarray}

Now, let us consider the projected spectral function,
\begin{eqnarray}
{\rho_\pm}(p_0)={1\over4}{\rm Tr}\left[P_\pm{\rho}(p_0)\right],\quad
{\rho_\pm}^{\rm OPE}(p_0)={1\over4}{\rm Tr}\left[P_\pm{\rho}^{\rm OPE}(p_0)\right],
\end{eqnarray}
where ${\rho_\pm}(p_0)={\rho_\mp}(-p_0)$
and ${\rho_\pm}^{\rm OPE}(p_0)={\rho_\mp}^{\rm OPE}(-p_0)$.
We approximate each of the positive and negative-parity contributions in the projected spectral function by the lowest pole plus continuum ansatz.
Namely, we parameterize the projected spectral function as,
\begin{eqnarray}\label{Phen}
\rho_\pm(p_0)&=&|\lambda_\pm|^2\delta(p_0-m_\pm)+|\lambda_\mp|^2\delta(p_0+m_\mp)
+[\theta(p_0-\omega_\pm)+\theta(-p_0-\omega_\mp)]\rho^{\rm OPE}_\pm(p_0),
\end{eqnarray}
where $\omega_+$ and $\omega_-$ denote the effective continuum thresholds for positive and negative-parity channels, respectively.
Therefore, the anzatz of Eq.~(\ref{Phen}) is expected to approximate the physical spectrum better than the usual lowest pole plus continuum model where only the lowest energy state of either positive or negative parity is taken into account as a pole and the rest is included in the common continuum term.
Substituting Eq.~(\ref{Phen}) to the right-hand side in Eq.~(\ref{QSR}) and using the Borel weight, $W(p_0)=p_0^n\exp(-p_0^2/M^2)$, we obtain the Borel sum rule,
\begin{eqnarray}\label{BSR}
\Pi_\pm^n(M,\omega_+,\omega_-)&\equiv&
\int_{-\omega_\mp}^{\omega_\pm}dp_0\rho_\pm^{\rm OPE}(p_0)p_0^n\exp(-{p_0^2\over M^2})
\cr&=&{m_\pm}^n|\lambda_\pm|^2\exp(-{{m_\pm}^2\over M^2})+(-m_\mp)^n|\lambda_\mp|^2\exp(-{{m_\mp}^2\over M^2}),
\end{eqnarray}
where the parameter of the weight function, $M$, is called the Borel mass.
The projected correlation function defined by Eq.~(\ref{BSR}) satisfies a relation,\begin{eqnarray}\label{Relation}
\Pi_\pm^n(M,\omega_+,\omega_-)=(-1)^n\Pi_\mp^n(M,\omega_+,\omega_-).
\end{eqnarray}
It should be noted that in Eq.~(\ref{BSR}) the positive and negative-parity baryons couple with each other.
Combining $\Pi_+^n$ ($\Pi_-^n$) defined by Eq.~(\ref{BSR}) with different $n$, we can eliminate $\lambda_\pm$ and express $m_+$ ($m_-$) by $m_-$ ($m_+$) as
\begin{eqnarray}\label{SRone}
m_+
&=&{m_-\Pi_+^{k+1}(M,\omega_+,\omega_-)+\Pi_+^{k+2}(M,\omega_+,\omega_-)\over
m_-\Pi_+^k(M,\omega_+,\omega_-)+\Pi_+^{k+1}(M,\omega_+,\omega_-)},
\cr
m_-
&=&{m_+\Pi_-^{l+1}(M,\omega_+,\omega_-)+\Pi_-^{l+2}(M,\omega_+,\omega_-)\over
m_+\Pi_-^l(M,\omega_+,\omega_-)+\Pi_-^{l+1}(M,\omega_+,\omega_-)}.
\end{eqnarray}
Taking $l\not=k$ in Eq.~(\ref{SRone}), we can solve for $m_\pm$ as
\begin{eqnarray}\label{SRtwo}
m_\pm&=&\Bigg[\sqrt{(\Pi_+^k\Pi_-^{l+2}-\Pi_+^{k+2}\Pi_-^l)^2
 +4(\Pi_+^k\Pi_-^{l+1}+\Pi_+^{k+1}\Pi_-^l)(\Pi_+^{k+1}\Pi_-^{l+2}+\Pi_+^{k+2}\Pi_-^{l+1})}
\cr&&\mp(\Pi_+^k\Pi_-^{l+2}-\Pi_+^{k+2}\Pi_-^l)\Bigg]\left/2(\Pi_+^k\Pi_-^{l+1}+\Pi_+^{k+1}\Pi_-^l)\right.\qquad(l\not=k).
\end{eqnarray}
For the ansatz Eq.~(\ref{Phen}) to be reasonable, Eq.~(\ref{BSR}) must satisfy the condition,
\begin{eqnarray}\label{Condition}
\Pi_\pm^{2n}(M,\omega_+,\omega_-)>0,
\end{eqnarray}
since $|\lambda_\pm|^2$ must be positive.
 
Comments on the difference of the present sum rule and the previous sum rules are in order here.
The authors of Ref.~\cite{chungII} studied the positive and negative-parity baryons in the QCD sum rule.
They derived sum rules using the dispersion relation in the variable $p^2$ as
\begin{eqnarray}\label{Dispersion}
\Pi_{i}(p^2)=\int_0^\infty{\rho_{i}(s)\over p^2-s}ds,
\end{eqnarray}
where $\Pi(p)=\p\Pi_1(p^2)+\Pi_2(p^2)$.
Then the spectral functions are given by
\begin{eqnarray}\label{PhenT}
\rho_1(p^2)&=&
\sum_n\left\{|\lambda^n_+|^2\delta\left(p^2-{m^n_+}^2\right)+|\lambda^n_-|^2\delta\left(p^2-{m^n_-}^2\right)\right\},
\cr
\rho_2(p^2)&=&
\sum_n\left\{|\lambda^n_+|^2m^n_+\delta\left(p^2-{m^n_+}^2\right)-|\lambda^n_-|^2m^n_-\delta\left(p^2-{m^n_-}^2\right)\right\}.
\end{eqnarray}
In order to extract the mass of the positive(negative)-parity baryon, they constructed the interpolating field in such a way that the contribution of $|\lambda^n_+|^2$ ($|\lambda^n_-|^2$) becomes as large as possible and just neglected the contribution from negative(positive)-parity baryon.
Therefore, the neglected contribution of the negative(positive)-parity baryon might contaminate the sum for the positive(negative)-parity baryon.
In particular, the sum rule for the negative-parity baryon is dangerous because the energy of the lowest negative-parity state is expected to be above the threshold of the positive-parity continuum.
Actually, they did not find a Borel stability in their sum rule for the mass of the negative-parity nucleon.
On the other hand, in the present sum rule we approximate each of the positive and negative-parity contributions in the projected spectral function by the lowest pole plus continuum ansatz.
Therefore, we can separately control the continuum thresholds of the positive and negative-parity states so that the assumed spectral function is expected to a better approximation to the physical spectrum.
The authors of Ref.~\cite{jido} proposed to use the \lq\lq old-fashioned'' correlation function in order to separate the terms of negative-parity baryons from those of positive-parity baryons by operating the projection operators.
Here, we would like to point out the difficulty in deriving sum rules using the \lq\lq old-fashioned''  correlation function.
The spectral function for the \lq\lq old-fashioned'' correlation function is given by
\begin{eqnarray}\label{OldSpectral}
\rho_{old}(p_0)=
P_+\sum_n|\lambda^n_+|^2\delta(p_0-m^n_+)+P_-\sum_n|\lambda^n_-|^2\delta(p_0-m^n_-),
\end{eqnarray}
which is obviously related to $\rho(p_0)$, Eq.~(\ref{Spectral}), as $\rho_{old}(p_0)=\theta(p_0)\rho(p_0)$.
Formally, the OPE of the spectral functions are given for the time-ordered correlation function by
\begin{eqnarray}\label{Spectral_OPE}
\rho^{OPE}(p_0)&=&
P_+\sum_n\left\{\left(|\lambda^n_+|^2+|\lambda^n_-|^2\right)\delta(p_0)+\left(-|\lambda^n_+|^2m^n_++|\lambda^n_-|^2m^n_-\right)\delta'(p_0)+\cdots\right\}
\cr&&+P_-\sum_n\left\{\left(|\lambda^n_+|^2+|\lambda^n_-|^2\right)\delta(p_0)+\left(|\lambda^n_+|^2m^n_+-|\lambda^n_-|^2m^n_-\right)\delta'(p_0)+\cdots\right\},
\end{eqnarray}
and for the \lq\lq old-fashioned'' correlation function by
\begin{eqnarray}\label{OldSpectral_OPE}
\rho^{OPE}_{old}(p_0)&=&
P_+\sum_n\left\{|\lambda^n_+|^2\delta(p_0)-|\lambda^n_+|^2m^n_+\delta'(p_0)+\cdots\right\}
\cr&&+P_-\sum_n\left\{|\lambda^n_-|^2\delta(p_0)-|\lambda^n_-|^2m^n_-\delta'(p_0)+\cdots\right\},
\end{eqnarray}
(It should be noted here that we exchanged the expansion in $1/p_0$ and the sum over poles.
That is why we obtain only terms of the delta function and its derivatives but not of the theta function.
But in the following discussion the existence of the delta function and its derivatives is essential and the lack of the theta function is unimportant.)
Clearly, one cannot obtain $\rho^{OPE}_{old}(p_0)$ from $\rho^{OPE}(p_0)$ by the relation, $\rho_{old}(p_0)=\theta(p_0)\rho(p_0)$. 
The terms, which include the delta function or its derivatives, do not keep track on whether they are from the positive energy part or the negative energy part, while the terms with $\theta(p_0)$ and $\theta(-p_0)$ obviously originate from the positive energy part and the negative energy part of the spectral function, respectively.
Thus, there is no way to determine the OPE of the \lq\lq old fashioned'' correlation function from that of the time ordered correlation function.

\section{Application to nucleon channel}

The general expression of the nucleon interpolating field without derivatives~\cite{ioffe,espriu,ioffeII} is given by
\begin{eqnarray}\label{NIF}
\eta_N=\epsilon_{abc}[(u_aCd_b)\g_5u_c+t(u_aC\g_5d_b)u_c],
\end{eqnarray}
where $u$ and $d$ are field operators of up and down quarks, respectively, $C$ denotes the charge conjugation operator and $a$, $b$ and $c$ are color indices. In Eq.~(\ref{NIF}), $t$ can be considered the tangent of a mixing angle and is called the mixing parameter in the following.
In the OPE of the nucleon correlation function, Wilson coefficients have been calculated in Ref.~\cite{espriu} for all the operators up to dimension~4 and the four-quark operator of dimension~6.
The coefficient for the quark-gluon mixed operator of dimension~5 has also been calculated in Ref~\cite{jido}. 
The OPE of the nucleon spectral function including these operators is given by
\begin{eqnarray}\label{Nrho}
\rho(p_0)&=&\gamma_0\{{5+2t+5t^2\over2^{11}\pi^4}p_0^5[\theta(p_0)-\theta(-p_0)]
\cr&&+{5+2t+5t^2\over2^{10}\pi^2}\GG p_0[\theta(p_0)-\theta(-p_0)]
\cr&&+{7t^2-2t-5\over24}\qq^2\delta(p_0)\}
\cr&&-{7t^2-2t-5\over64\pi^2}p_0^2\qq[\theta(p_0)-\theta(-p_0)]
\cr&&+{3(t^2-1)\over64\pi^2}\qGq[\theta(p_0)-\theta(-p_0)],
\end{eqnarray}
where we adopt the factorization hypothesis for the four-quark condensate. 
In Eq.~(\ref{Nrho}), $\alpha_s=g^2/4\pi$ with $g$ being the strong coupling constant and $G_{\mu\nu}= G_{\mu\nu}^a\lambda^a/2$ where $G_{\mu\nu}^a$ is the gluon field tensor and $\lambda^a$ is the usual Gell-Mann SU(3) matrix.

\begin{figure}
  \begin{center}
    \leavevmode
  \epsfxsize=4.5 in
  \epsfbox{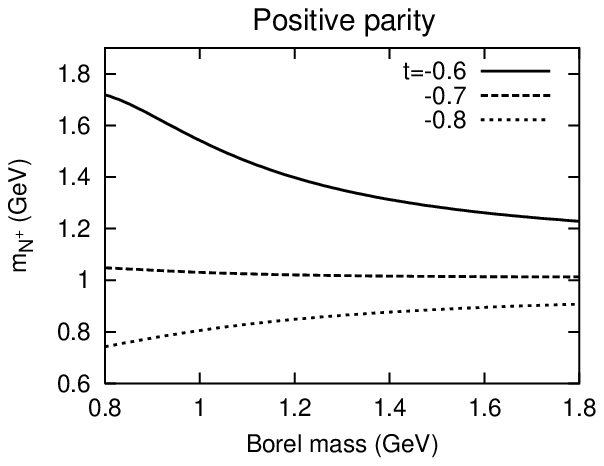}
  \end{center}
\caption{
Mass of the positive-parity nucleon as a function of Borel mass with the mixing parameters $t=-0.6$ (solid line), $-0.7$ (dashed line) and $-0.9$ (dotted line).}
  \label{figI}
\end{figure}

\begin{figure}
  \begin{center}
    \leavevmode
  \epsfxsize=4.5 in
  \epsfbox{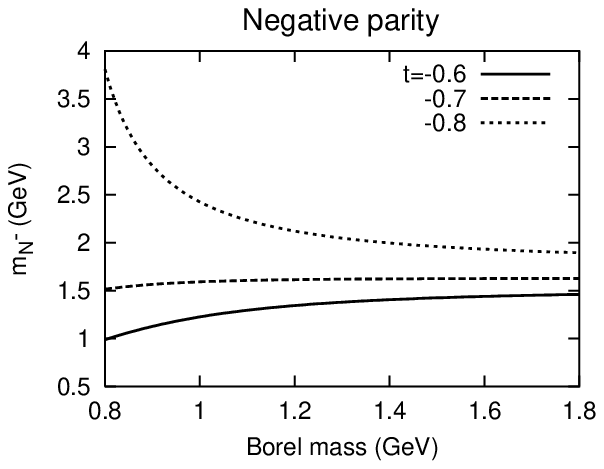}
  \end{center}
\caption{
Mass of the negative-parity nucleon as a function of Borel mass with the mixing parameters $t=-0.6$ (solid line), $-0.7$ (dashed line) and $-0.9$ (dotted line).
}
  \label{figII}
\end{figure}

Now we study the predictability of the present sum rule.
Substituting Eq.~(\ref{Nrho}) into the integral in Eq.~(\ref{BSR}) we derive the Borel sum rules for the positive and negative-parity nucleons.
In the following demonstration, we choose $\omega_+=1.44$~GeV and $\omega_-=1.65$~GeV for the effective continuum threshold which correspond to the masses of $N(1440)$~$P_{11}$ and $N(1650)$~$S_{11}$ resonance states, respectively.
For the QCD parameters we choose the standard values:
\begin{eqnarray}\label{Parameter}
\qq=-(0.23\;{\rm GeV})^3,\quad \GG=(0.33\;{\rm GeV})^4,\quad m_0^2\equiv{\qGq\over\qq}=0.8\;{\rm GeV}^2.
\end{eqnarray}
The gluon condensate is extracted from the charmonium sum rules in Ref.~\cite{SVVZ} and $m_0^2$ from the baryon sum rules in Refs.~\cite{BandI,OandP}.
We find the optimal choice of the Borel weight, $k$ and $l$ in Eq.~(\ref{SRtwo}), to be $l=0$ and $k=2$ ($l=2$ and $k=0$) by investigating the Borel stability of the sum rules for $0\leq k,l\leq5$.
Finally we obtain
\begin{eqnarray}\label{SRfinal}
m_\pm={\sqrt{((\Pi_+^2)^2-\Pi_+^0\Pi_+^4)^2
-4(\Pi_+^1\Pi_+^2-\Pi_+^0\Pi_+^3)(\Pi_+^2\Pi_+^3-\Pi_+^1\Pi_+^4)}
\pm((\Pi_+^2)^2-\Pi_+^0\Pi_+^4)
\over2(\Pi_+^1\Pi_+^2-\Pi_+^0\Pi_+^3)},
\end{eqnarray}
where we use the relation (\ref{Relation}).
From Eq.~(\ref{SRfinal}) we determine the masses of the positive and negative-parity nucleons.
The mixing parameter, $t$, is determined by the following procedure.
From the condition (\ref{Condition}) we restrict the mixing parameter to be $t<-0.5$ or $0.6<t$ in the region of $M>0.8$~GeV.
Performing the Borel stability analysis in the region of $0.8\;{\rm GeV}<M<1.8\;{\rm GeV}$, we find the optimal value for the mixing parameter.
Figures~\ref{figI} and \ref{figII} display how sensitive the Borel stability is to the mixing parameter.
Figs.~\ref{figI} and \ref{figII} show $m_{N^+}$ and $m_{N^-}$, respectively,  vs. the Borel mass for $t=-0.6$, $-0.7$ and $-0.8$. 
One sees that the Borel stability is sensitive to the mixing parameter and that the most stable plateau as a function of the Borel mass appears when $t=-0.7$ both for $m_{N^+}$ and $m_{N^-}$ have.
For this optimal choice of the mixing parameter the positive and negative-parity nucleon masses vary little in the region from $M\approx0.8$~GeV to $1.8$~GeV.
Actually, varying the Borel mass from $0.8$~GeV to $1.2$~GeV (from $M=1.4$~GeV to $1.8$~GeV) changes $m_{N^+}$ ($m_{N^-}$) by less than 3\% (1\%).
It should be noted that there appears a rather stable plateau also for $t\sim0.9$ but the stability is worse than that for $t\sim-0.7$.
From the Borel curves with $t=-0.7$ we obtain the masses of the positive and negative-parity nucleons as
\begin{eqnarray}
m_{N^+}=1.0\;{\rm GeV},\quad m_{N^-}=1.6\;{\rm GeV}.
\end{eqnarray}
The calculated values are a bit greater than the experimental values, $m_{N^+}=0.939\;{\rm GeV}$
and $m_{N^-}=1.535\;{\rm GeV}$.
It is, however, remarkable that the calculated mass difference, $m_{N^-}-m_{N^+}=0.6\;{\rm GeV}$, is very close to the experimental one, $0.596$~GeV.
The uncertainties of $\qq$, $\GG$ and $m_0^2$ are 40\%, 30\% and 10\%, respectively.
These errors change the positive and negative-parity nucleon masses by $\pm 0.13$~GeV and  $\pm 0.10$~GeV, respectively.

Now, we check how sensitive the calculated masses are to the effective continuum thresholds.
Varying $\omega_+$ from 1.26~GeV to 1.62~GeV, which corresponds to shifting the Roper resonance mass by the Breit-Wigner width, changes  the positive and negative-parity nucleon masses by $\pm0.04$~GeV and $\pm0.32$~GeV, respectively.
Similarly, varying $\omega_-$ from 1.57~GeV to 1.73~GeV, which corresponds to shifting the $N(1650)$~$S_{11}$ resonance mass by the width, changes them by less than a few percent.
Thus we find that the negative-parity nucleon mass is very sensitive to the continuum threshold, $\omega_+$, while the positive-parity nucleon mass is not so sensitive to the continuum thresholds.
This is natural because the positive-parity nucleon is the lowest of all positive and negative-parity states, while the negative-parity nucleon is above the positive-parity continuum threshold.

\section{Summary}
We have proposed a new approach of the QCD sum rule in which positive and negative-parity baryons couple with each other.
Since the baryon interpolating field couples to negative as well as positive-parity baryon states, the time-ordered correlation function include both contributions.
Explicitly taking account of positive and negative-parity states, we have derived sum rules by means of the dispersion relation in energy.
Then, we have applied the method to the nucleon channel and investigated the parity splitting of the nucleons.
The obtained sum rules turn out to have a very good Borel stability.
This implies that the ansatz for the spectral function in the present sum rule approximates the physical spectrum better than the usual lowest pole plus continuum ansatz.
The predicted masses of the positive and negative nucleons reproduce the experimental ones fairly well. 
Especially, the mass difference is extremely close to the experimental value.

It is well known that for the vector meson states the application of the QCD sum rule has been quite successful where the Borel curves have almost perfectly stable plateaus~\cite{SVZ,RRY}.
In the sum rule for the nucleon mass, however, the stability of the Borel curve is not so good when we take a simple lowest pole plus continuum ansatz for the spectral function.
In the present approach once we take positive and negative-parity states explicitly into account,  the Borel stability has been drastically improved not only for the positive-parity nucleon but also for the negative-parity nucleon as we have demonstrated in Sec.~III.
We should, however, mention that another possibility of improving the Borel stability has been previously proposed by Dorokhov and Kochelev~\cite{DandK} and Forkel and Banerjee~\cite{HandB} in which they have taken account of the effects of direct instantons in the correlation functions.

The present results suggest that the parity splitting of baryons can be studied in the framework of the QCD sum rule once the coupling of positive and negative-parity states is appropriately taken into account.
Therefore, it is a future problem to apply the present approach to the parity splitting of various baryons.
In particular, it is extremely interesting to investigate the pentaquark baryon since its parity, which is not experimentally determined, is crucial to understand the existence of the pentaquark baryon.

%%%%%%%%%%%%%%%%%%%%%%%%%%%%%%%%%%%%%%%%%%%%%%%
\def\Ref#1{[\ref{#1}]}
\def\Refs#1#2{[\ref{#1},\ref{#2}]}
\def\npb#1#2#3{{Nucl. Phys.\,}{\bf B{#1}}\,(#3)\,#2}
\def\npa#1#2#3{{Nucl. Phys.\,}{\bf A{#1}}\,(#3)\,#2}
\def\np#1#2#3{{Nucl. Phys.\,}{\bf{#1}}\,(#3)\,#2}
\def\plb#1#2#3{{Phys. Lett.\,}{\bf B{#1}}\,(#3)\,#2}
\def\prl#1#2#3{{Phys. Rev. Lett.\,}{\bf{#1}}\,(#3)\,#2}
\def\prd#1#2#3{{Phys. Rev.\,}{\bf D{#1}}\,(#3)\,#2}
\def\prc#1#2#3{{Phys. Rev.\,}{\bf C{#1}}\,(#3)\,#2}
\def\prb#1#2#3{{Phys. Rev.\,}{\bf B{#1}}\,(#3)\,#2}
\def\pr#1#2#3{{Phys. Rev.\,}{\bf{#1}}\,(#3)\,#2}
\def\ap#1#2#3{{Ann. Phys.\,}{\bf{#1}}\,(#3)\,#2}
\def\prep#1#2#3{{Phys. Reports\,}{\bf{#1}}\,(#3)\,#2}
\def\rmp#1#2#3{{Rev. Mod. Phys.\,}{\bf{#1}}\,(#3)\,#2}
\def\cmp#1#2#3{{Comm. Math. Phys.\,}{\bf{#1}}\,(#3)\,#2}
\def\ptp#1#2#3{{Prog. Theor. Phys.\,}{\bf{#1}}\,(#3)\,#2}
\def\ib#1#2#3{{\it ibid.\,}{\bf{#1}}\,(#3)\,#2}
\def\zsc#1#2#3{{Z. Phys. \,}{\bf C{#1}}\,(#3)\,#2}
\def\zsa#1#2#3{{Z. Phys. \,}{\bf A{#1}}\,(#3)\,#2}
\def\intj#1#2#3{{Int. J. Mod. Phys.\,}{\bf A{#1}}\,(#3)\,#2}
\def\sjnp#1#2#3{{Sov. J. Nucl. Phys.\,}{\bf #1}\,(#3)\,#2}
\def\jtep#1#2#3{{Sov. Phys. JTEP\,}{\bf #1}\,(#3)\,#2}
\def\pan#1#2#3{{Phys. Atom. Nucl.\,}{\bf #1}\,(#3)\,#2}
\def\app#1#2#3{{Acta. Phys. Pol.\,}{\bf #1}\,(#3)\,#2}
\def\jmp#1#2#3{{J. Math. Phys.\,}{\bf {#1}}\,(#3)\,#2}
\def\cp#1#2#3{{Coll. Phen.\,}{\bf {#1}}\,(#3)\,#2}
\def\epjc#1#2#3{{Eur. Phys. J.\,}{\bf C{#1}}\,(#3)\,#2}
\def\zetf#1#2#3{{Zh. Ekps. Teor. Fiz.\,}{\bf #1}\,(#3)\,#2}
\def\yf#1#2#3{{Yad. Fiz.\,}{\bf A{#1}}\,(#3)\,#2}

%%%%%%%%%%%%%%%%%%%%%%%%%%%%%%%%%%%%%%%%%%%%%%%%%%%%%%%%%%%%%%%%%%
%%%%%%%%%%%%%%%%%%%%%%%%%%%%%%%%%%%%%%%%%%%%%%%%%%%%%%%%%%%%%%%%%%
%%%%%%%%%%%%%%%%%%%%%%%%%%%%%%%%%%%%%%%%%%%%%%%%%%%%%%%%%%%%%%%%%%

\end{document}